\documentclass[11pt,a4paper]{article}
\usepackage{jheppub}

\usepackage{bm}
\usepackage{bbm}

\DeclareMathOperator{\tr}{Tr}
\newcommand{\imag}{i}
\newcommand{\dd}{\mathrm{d}}
\newcommand{\openone}{\mathbbm{1}}
\newcommand{\vek}[1]{\bm{#1}}
\newcommand{\norvec}[1]{|\vec{#1}|}
\newcommand{\Z}{\mathcal{Z}}
\newcommand{\Nf}{N_{\text f}}

\title{Two-color QCD via dimensional reduction}

\author[a]{Tian Zhang,}
\author[b,c]{Tom\'{a}\v{s} Brauner,}
\author[d]{Aleksi Kurkela}
\author[b]{and Aleksi Vuorinen}

\affiliation[a]{Institute for Theoretical Physics, Goethe University, Frankfurt am Main, Germany}
\affiliation[b]{Faculty of Physics, University of Bielefeld, Bielefeld, Germany}
\affiliation[c]{Department of Theoretical Physics, Nuclear Physics Institute ASCR, \v{R}e\v{z}, Czech Republic}
\affiliation[d]{Department of Physics, McGill University, Montr\'eal, Canada}

\emailAdd{tzhang@th.physik.uni-frankfurt.de}
\emailAdd{tbrauner@physik.uni-bielefeld.de}
\emailAdd{kurkela@physics.mcgill.ca}
\emailAdd{vuorinen@physik.uni-bielefeld.de}

\abstract{We study the thermodynamics of two-color QCD at high temperature and/or density using a dimensionally reduced superrenormalizable effective theory, formulated in terms of a coarse grained Wilson line. In the absence of quarks, the theory is required to respect the Z(2) center symmetry, while the effects of quarks of arbitrary masses and chemical potentials are introduced via soft Z(2) breaking operators. Perturbative matching of the effective theory parameters to the full theory is carried out explicitly, and it is argued how the new theory can be used to explore the phase diagram of two-color QCD.}


\begin{document}
\maketitle

\section{Introduction}
\label{sec:intro}

Dimensionally reduced effective theories are known to provide a physically transparent and computationally efficient way to access the long distance dynamics of strongly interacting matter at high temperatures~\cite{Appelquist:1981vg}. Their construction requires the existence of a clear scale hierarchy between the hard scale $\pi T$ and the soft scale $gT$ (as well as the magnetic scale $g^2T$), where $T$ is the temperature and $g$ the gauge coupling (cf.~the discussion and references in ref.~\cite{Vuorinen:2006nz}). At temperatures far above the deconfinement transition, this is indeed guaranteed by asymptotic freedom, but there are indications that at least a modest hierarchy exists all the way down to the phase transition region~\cite{Kaczmarek:1999mm}. A lot of effort has subsequently been devoted to the study of Electrostatic QCD (EQCD)~\cite{Kajantie:1995dw,Braaten:1995jr,Kajantie:1997tt}, both perturbatively and using lattice simulations (see e.g.~refs.~\cite{Kajantie:1998yc,Kajantie:2003ax}).

Despite the above, it is commonly believed that dimensionally reduced theories have little or nothing to say about the deconfinement phase transition of QCD. This is to a large extent due to the fact that the construction of EQCD explicitly breaks the center symmetry, which is known to play an important role in the phase transition dynamics even with dynamical fermions present~\cite{Pisarski:2006hz}. For the case of pure Yang-Mills theory, this issue was addressed in refs.~\cite{Vuorinen:2006nz,Kurkela:2007dh,deForcrand:2008aw}, where a new type of center-symmetric dimensionally reduced effective theory (ZQCD) was proposed. Such an effective theory is expected to provide a quantitatively reliable description of QCD thermodynamics down to temperatures close to the deconfinement transition. A key ingredient in this work was the replacement of the temporal gauge field of EQCD by a coarse grained Wilson line, at the same time retaining the superrenormalizability of the theory (for closely related work, see also refs.~\cite{Pisarski:2006hz,Wozar:2007tz,Fromm:2011qi}). In particular, ref.~\cite{deForcrand:2008aw} demonstrated that upon perturbatively matching two-color ZQCD to the full theory at high temperatures, one obtains a consistent effective theory for SU(2) Yang-Mills theory, exhibiting --- as expected --- a second order confining phase transition. It was further found that the predictions of the theory are to a very good accuracy independent of the value of the single hard (order $g^0T$) parameter that was introduced in the coarse graining procedure.

The success of ZQCD as an effective theory for high temperature Yang-Mills theory naturally raises the question, whether similar methods can be used to build an effective theory for full QCD with dynamical quarks, despite the fact that the latter explicitly break the center symmetry. While this type of an approach would naturally not capture the dynamics relevant for the breaking of chiral symmetry, it should serve as a useful starting point for studying the thermodynamics of QCD \textit{at least} when the quark masses are sufficiently large. Constructing and studying such a theory for the case of two-color QCD is the main objective of this paper.

Alongside with the inclusion of fermions, we will in this article discuss another important issue related to the original construction of ZQCD in refs.~\cite{Vuorinen:2006nz,deForcrand:2008aw}. As we will explicitly show in the following sections, these original works contain an identical algebraic error that affects the perturbative matching of the effective theory parameters. In particular, the domain wall tension and width that were previously used to determine two of the effective theory parameters will be shown to be uniquely determined by the perturbative matching of the ZQCD Lagrangian to EQCD, and thus become genuine predictions of the theory.\footnote{While we show this  explicitly only for gauge group SU(2), the SU(3) case follows in complete analogy.} This discovery has little significance for the qualitative findings of refs.~\cite{Vuorinen:2006nz,deForcrand:2008aw}, yet it has important implications for how the matching process should be carried out.

The paper is organized as follows. In section~\ref{sec:EFT}, we remind the reader of the construction of center-symmetric effective theories \`a la refs.~\cite{Vuorinen:2006nz,deForcrand:2008aw}, after which we discuss the new Z(2) breaking operators that emerge upon inclusion of fundamental fermions in the case of two colors. In section~\ref{sec:fields}, we then identify the degrees of freedom of the effective theory and explain how they are related to those of EQCD, while in section~\ref{sec:matching}, the detailed matching of the ZQCD parameters to the full theory is carried out. Section~\ref{sec:solitons} finally contains the first predictions of the new theory, followed by a concluding section~\ref{sec:conclusions}, in which we discuss the prospects of using the effective theory as a tool for studying the thermodynamics of the full one. Some technical details are left to the appendices; as a particularly useful byproduct of our calculations, we will in appendix~\ref{app:EQCD} present analytic expressions for all $N_c=2$ EQCD parameters for general values of the temperature, quark masses and chemical potentials.


\section{The effective theory}
\label{sec:EFT}

Our purpose in this paper is to formulate a three-dimensional superrenormalizable effective theory for two-color QCD, valid for high temperatures and finite quark chemical potentials. We require that in the absence of fermions the theory respect the Z(2) center symmetry, and that the effects of fermions can be incorporated through Z(2) breaking operators that are soft, i.e.~suppressed by some power of the gauge coupling. To this end, we follow the path laid in refs.~\cite{Vuorinen:2006nz,deForcrand:2008aw} and introduce as our degrees of freedom a coarse grained Wilson line $\Z(\vek x)$ and a magnetic gluon field $\vek A(\vek{x})$. The gauge transformation acts on these fields as
\begin{equation}
\Z(\vek{x})\to s(\vek{x})\Z(\vek{x})s(\vek{x})^\dagger,\qquad\vek A(\vek{x})\to s(\vek{x})[\vek A(\vek{x})+\imag\vek\nabla]s(\vek{x})^\dagger,
\end{equation}
where $s(\vek{x})\in$ SU(2), while the Z(2) transformation acts on the $\Z$ field as
\begin{equation}
\Z(\vek{x})\to{\rm e}^{\imag\pi n}\Z(\vek{x})=\pm\Z(\vek{x}).
\label{center}
\end{equation}
The relation of the latter transformation to the SU(2) gauge invariance of the full theory is explained in detail in appendix~\ref{app:center}.

As discussed at length in ref.~\cite{deForcrand:2008aw}, a unique feature of the gauge group SU(2) is that the coarse graining procedure almost preserves the group property of the Wilson line, as an arbitrary sum of SU(2) matrices is itself an SU(2) matrix up to a multiplicative real factor. This implies that we may parameterize the field $\Z$ in the form
\begin{equation}
\Z=\frac{1}{2}\Bigl(\Sigma\openone+\imag\vec\Pi\cdot\vec\sigma\Bigr),
\label{Zdef}
\end{equation}
where $\vec\Pi\cdot\vec\sigma\equiv\Pi_a\sigma_a$ ($\sigma_a$ denoting the Pauli matrices), and $\Sigma$ and $\Pi_a$, $a=1,2,3$, are real scalar fields. Out of these four degrees of freedom we expect three to correspond to the light adjoint Higgs field of EQCD, while one should be an unphysical auxiliary field that effectively decouples from the dynamics of the light fields and has a mass of the order of the cutoff scale of the effective theory ($\sim T$), corresponding to the inverse length scale introduced by the coarse graining. While the heavy auxiliary field decouples from the dynamics in the infrared, its fluctuations in the ultraviolet render the theory superrenormalizable, providing important technical simplifications.

To obtain the Lagrangian density of the effective theory, we collect all superrenormalizable operators up to fourth order in the fields that respect three-dimensional gauge invariance.\footnote{While in a three-dimensional theory operators of order five in the fields are in principle still relevant and those of order six marginal, we exclude them from our consideration, as their contributions to physical quantities are suppressed by (large) positive powers of the ratio of the effective and full theory energy scales.} This leads to the expression
\begin{equation}
\mathcal{L}=\frac1{g_3^2}\left[\frac12\tr F_{ij}^2+\tr\Bigl(D_i \Z^{\dagger}D_i\Z\Bigr)+V(\Z)\right],
\label{lageff2}
\end{equation}
where $g_3$ is the effective theory gauge coupling, $D_i\equiv\partial_i-\imag[A_i,\,\cdot\,]$, $F_{ij} \equiv \partial_iA_j-\partial_jA_i-\imag[A_i,A_j]$, and the potential $V(\Z)$ reads
\begin{equation}
V(\Z)=b_1\Sigma^2+b_2\vec\Pi^2+c_1\Sigma^4+c_2(\vec\Pi^2)^2+c_3\Sigma^2\vec\Pi^2+d_1\Sigma^3+d_2\Sigma\vec\Pi^2.
\end{equation}
Here, all terms with the exception of the last two operators in the potential respect the Z(2) center symmetry, and were present in the model constructed for pure SU(2) Yang-Mills theory in ref.~\cite{deForcrand:2008aw}; the Z(2) violating operators then clearly result from the presence of quarks. It is a straightforward exercise to verify that eq.~\eqref{lageff2} really is the most general Lagrangian compatible with the required symmetries: simple redefinitions of the fields allow us to combine independent kinetic terms for $\Sigma$ and $\vec\Pi$ and remove a Z(2) breaking term linear in $\Sigma$, while a term cubic in $\vec\Pi$ is forbidden by the vanishing of the symmetric structure constant $d_{abc}$ in SU(2).

Next, we follow the rationale of refs.~\cite{Vuorinen:2006nz,deForcrand:2008aw} and split the effective theory potential into `hard' and `soft' parts, parameterized respectively by the ${\mathcal O}(g^0)$ constants $h_i$ and $s_i$. This results in the alternative expression
\begin{equation}
V(\Z)=h_1\tr(\Z^\dagger \Z)+h_2(\tr \Z^\dagger \Z)^2+g_3^2\left[\frac{s_1}2\vec\Pi^2+\frac{s_2}4(\vec\Pi^2)^2+s_3\Sigma^4+\frac{s_4}{2}\Sigma^3+\frac{s_5}2\Sigma\vec\Pi^2 \right],
\label{pot}
\end{equation}
where we have assumed the Z(2) breaking couplings $d_i$ to be of the soft type. The kinetic terms as well as the hard part of the potential possess an extended (global) SU(2)$\times$SU(2) invariance,
\begin{equation}
\Z\to\Omega_1\Z\Omega_2,\qquad \Omega_i\in\mathrm{SU(2)},
\end{equation}
which will later be seen to translate to a shift invariance of the light physical fields of ZQCD upon integrating the heavy one out. As a consequence, the hard part of the potential is minimized by all matrices that are special unitary up to a common real factor. It is only the soft terms that provide the ${\mathcal O}(g^2)$ structure inside this `valley', necessary to match the effective theory potential to that of the full theory Wilson line, cf.~eq.~\eqref{match_V}. For later use, we note that comparing the two sets of coupling constants produces the relations
\begin{equation}
\begin{split}
b_1&=\frac{1}{2}h_1,\qquad b_2=\frac{1}{2}(h_1+g_3^2s_1),\\
c_1&=\frac{1}{4}h_2+g_3^2s_3,\qquad c_2=\frac{1}{4}(h_2+g_3^2s_2),\qquad c_3=\frac{1}{2}h_2,\\
d_1&=\frac{1}{2}g_3^2s_4,\qquad d_2=\frac{1}{2}g_3^2s_5.
\end{split}
\label{bd1}
\end{equation}


\section{Identification of the fields}
\label{sec:fields}

It is well known that at high temperatures, where the renormalized gauge coupling becomes small, the
Wilson line effectively freezes to the global minimum of its perturbative effective
potential~\cite{Gross:1980br,Weiss:1980rj,Weiss:1981ev} (for more recent work regarding the
effective potential of high-temperature QCD, see e.g.~ref.~\cite{Megias:2003ui}), and to correctly
describe its long distance dynamics it is sufficient to consider only its small fluctuations around
it. It is thus natural to require that the predictions of our effective theory reduce to those of
EQCD in the same limit, as the Lagrangian of EQCD can be obtained from an expansion of the Wilson
line potential in powers (and derivatives) of the temporal component of the gauge field, $A_0$. This
property can be most straightforwardly ensured by explicitly integrating out the heavy degree of
freedom in the vicinity of one of the minima of the ZQCD effective potential, and by matching the
resulting non-center-symmetric (even in the absence of fermions) theory to EQCD. Through this
procedure, the light field of ZQCD becomes associated with the adjoint Higgs $A_0$ of EQCD, and we
automatically obtain the values of several of the effective theory parameters.

In this and the following section, we will explicitly perform the high temperature matching of ZQCD to EQCD, and find the values of the $s_i$, i.e.~the soft parameters of ZQCD. We begin this by parameterizing the field $\Z$ as in eq.~(26) of ref.~\cite{deForcrand:2008aw},
\begin{equation}
\Z=\frac v2\openone+\frac{g_3}2(\phi\openone+\imag\vec\chi\cdot\vec\sigma),
\label{Zexppar}
\end{equation}
which amounts to the redefinition $\Sigma=v+g_3\phi$ and $\vec\Pi=g_3\vec\chi$, where $v$ is a real positive number, chosen so that $\langle\Z\rangle=(v/2)\openone$. Clearly, the precise choice of the parametrization of $\Z$ can have no effect on the physics, as long as it contains the correct degrees of freedom: once the effective theory is matched to the full theory properly, it will automatically reproduce the correct long distance physics. One should nevertheless note that, had we chosen to use a non-linear field parametrization, we would have had to consider the Jacobian associated with the change of variables in the defining path integral of the theory.

Upon rewriting eq.~\eqref{Zexppar} as $\Z=\frac{v+g_3\phi}{2}[\openone+\imag(g_3/v)\vec\chi\cdot\vec\sigma]+\mathcal O(g^2)$ and comparing with the full theory Wilson line,
\begin{equation}
\Omega(\vek{x})\equiv{\cal P}\exp\left[\imag g\int_0^{\beta}\dd\tau\,A_0(\tau,\vek{x})\right]=\openone+\imag g\int_0^{\beta}\dd\tau\,A_0(\tau,\vek{x}) +{\mathcal O}(g^2),
\label{omega}
\end{equation}
we identify the real scalar field $\phi$ as the auxiliary heavy degree of freedom of the effective theory (to leading order). Subsequently, we associate the field $\chi_a$ with the light, physical field that corresponds to the adjoint scalar $A_0$ of EQCD, which, together with the identification of the effective theory gauge coupling, $g_3^2=g^2T+ {\mathcal O}(g^4)$, fixes the leading order value of the parameter $v$,
\begin{equation}
v=2T+\mathcal O(g^2).
\label{v2T}
\end{equation}

Beyond the field identifications, the matching of ZQCD to the full theory is performed by demanding
that the long distance behavior of static gluonic correlators is correctly reproduced by the
effective theory, order by order in a weak coupling expansion. As the effects of the soft couplings
$s_i$ are suppressed by a factor $g_3^2$ in comparison with the hard ones $h_i$, one-loop graphs
with only hard vertices enter the effective theory calculation with the same power of $g_3$ as tree
graphs containing one soft vertex. This implies that to obtain the correlators in a consistent
manner, we need to determine the one-loop effective potential of ZQCD. This function can be read off
from eq.~(18) of ref.~\cite{deForcrand:2008aw} for the case of pure SU(2) Yang-Mills theory, and it
has been generalized to include the effects of fermionic operators in appendix~\ref{app:effpot} of
this paper. Inspecting the result reveals the anticipated effect that the degeneracy of the two
minima present in the center symmetric case is broken by the nonzero values of $s_{4,5}$. Without
lack of generality, we may choose $s_4<0$ so that the ground state expectation value of $\Sigma$ is
positive. Solving for the minimum of the potential iteratively, we find then
$v=v_0+g_3^2v_2+\dotsb$, where
\begin{equation}
v_0=\sqrt{-\frac{h_1}{h_2}},\qquad v_2=-\frac1{2h_2}\left(4s_3v_0+\frac32s_4\right)+\frac{3}{8\pi}\sqrt{2h_2}.
\end{equation}
Comparing this to the identification made in eq.~\eqref{v2T}, we infer from here the first nontrivial relation amongst our effective theory parameters,
\begin{equation}
h_1+4T^2h_2=0.
\label{match1}
\end{equation}


\section{Matching of the soft parameters}
\label{sec:matching}

In this section, we will perturbatively determine the values of the soft ZQCD parameters $s_i$. We begin this in section \ref{subsec:couplings} by integrating out the heavy field $\phi$ from the effective theory, requiring that the resulting Lagrangian for $\vec\chi$ agrees with that of EQCD. After this, we will in section \ref{subsec:wall} match the remaining soft effective theory parameters by demanding that the global structure of the one-loop effective potential of ZQCD agrees with that of the full theory.


\subsection{Perturbative matching of the Lagrangians}
\label{subsec:couplings}

Our first goal will be to explicitly integrate out the heavy auxiliary field $\phi$ in order to obtain an effective potential for $\vec\chi$ only, to be compared with the effective potential of the $A_0$ field in EQCD. At the level of the quantum (Wilsonian) effective action, integrating out a given field amounts to eliminating it using its equation of motion or, equivalently, adding to the action of the other fields all \emph{tree-level} Feynman graphs containing this field in the internal lines and all other fields as external legs.

At this point, the SU(2)$\times$SU(2) invariance of the hard part of the Lagrangian~\eqref{lageff2} proves its utility. The fields $\vec\chi$ can namely be identified as the Nambu-Goldstone bosons stemming from the spontaneous breaking of this extended symmetry by the nonzero expectation value $\langle\Z\rangle$. As a consequence, any contribution to the static correlators of $\vec\chi$ must come with at least one factor of $s_i$, and in particular, the one-loop part of the effective potential of ZQCD --- the second line of eq.~\eqref{VNLO} --- need not be taken into account. In addition, the gauge-fixing dependent part of the effective potential matches automatically to EQCD.

Keeping only terms up to fourth order in $\vec\chi$ and rescaling the spatial gluon field to achieve canonical normalization of its kinetic term, we arrive at a Lagrangian for this field that has the exact same form as that of EQCD,
\begin{equation}
\mathcal L_{\text{light}}=\frac12\tr F_{ij}^2+\frac12(D_i\vec\chi)^2+\frac12m_\chi^2\vec\chi^2+\frac{\tilde\lambda}8(\vec\chi^2)^2+\dotsb,
\end{equation}
with the mass parameter and quartic coupling reading
\begin{equation}
\begin{split}
m_\chi^2&=g_3^2\left(s_1-4s_3v_0^2-\frac32s_4v_0+s_5v_0\right),\\
\tilde\lambda&=2g_3^4\left(s_2+4s_3+\frac{3s_4}{4v_0}-\frac{s_5}{v_0}\right).
\end{split}
\label{ZQCDcoeffs}
\end{equation}
Note that when expressed in terms of Feynman diagrams, the quartic coupling $\tilde\lambda$ consists of two contributions, one from a soft operator of the $(\vec\chi^2)^2$ type, and another from a soft, SU(2)$\times$SU(2) breaking mass correction to a $\phi$ propagator connecting two hard cubic $\phi\vec\chi^2$ vertices. The latter contribution was missed in refs.~\cite{Vuorinen:2006nz,deForcrand:2008aw}, which resulted in an incorrect matching condition for the quartic coupling. The above expression for $\tilde\lambda$ thus corrects the corresponding expression in eq.~(32) of ref.~\cite{deForcrand:2008aw} and further generalizes it by including the effects of explicit Z(2) breaking due to dynamical quarks.\footnote{It should be additionally noted that the same, previously missed diagram with two cubic vertices leads to the generation of kinetic terms of the type $\vec\chi^2(D_i\vec\chi)^2$ and $(\vec\chi\cdot D_i\vec\chi)^2$ with couplings of order $\mathcal O(g_3^2)$, which enter the EQCD Lagrangian only at order $\mathcal O(g_3^4)$. Such terms can in principle be canceled by adding similar (non-renormalizable) operators to the effective theory Lagrangian of eq.~\eqref{lageff2}.}

Finally, the expressions in eq.~\eqref{ZQCDcoeffs} can be equated with their EQCD counterparts, given in eq.~\eqref{EQCDcoeffs}. This gives us two new matching conditions,
\begin{equation}
\begin{split}
s_1-4s_3v_0^2-\frac32s_4v_0+s_5v_0&=\frac{2T}3-\frac{T\kappa_0^-}{\pi^2},\\
2s_2+8s_3+\frac{3s_4}{2v_0}-\frac{2s_5}{v_0}&=\frac{2}{3\pi^2T}+\frac{\kappa_2^-}{12\pi^2T},
\end{split}
\label{match2}
\end{equation}
where the constants $\kappa_\ell^\pm$, parametrizing the effects of the quarks, are defined in eq.~\eqref{kappal}.


\subsection{The Z(2) breaking parameters}
\label{subsec:wall}

The matching conditions of eq.~\eqref{match2} should be viewed as fixing the values of two linear combinations of $s_{1,2,3}$ --- an interpretation that becomes trivial in the limit of unbroken center symmetry. In contrast, a third, independent linear combination of these parameters does not affect the physics of the soft scale at the leading order at all, and can thus take any value. This is because in the nonlinear version of our theory, where the heavy mode has been integrated out in a center-symmetric fashion, the three operators multiplying the coefficients $s_{1,2,3}$ are not independent, but there is a linear relation between $\Sigma^4$, $\vec\Pi^2$ and $(\vec\Pi^2)^2$. As the linear and nonlinear models describe the same long distance physics, there \emph{must} be one combination of $s_{1,2,3}$ that is left undetermined by the leading order matching of the linear theory, and can only be found through a higher order computation. The insensitivity of the long distance physics to this linear combination will be further demonstrated in section~\ref{sec:solitons}, where the domain wall solution of the field equations of motion is discussed.

As EQCD violates the center symmetry explicitly, it is clear that the parameters $s_{4,5}$, which facilitate the soft breaking of this symmetry in the presence of fermions, cannot be found by matching to EQCD. To determine their values, we instead have to consider the global structure of the ZQCD effective potential, which we do by applying the Nielsen theorem~\cite{Nielsen:1975fs} and concentrating on the second stationary point (local minimum) of the effective potential that provides additional gauge invariant observables. A natural measure of the center symmetry breaking is the energy density difference of the absolute and metastable minima, which on the full theory side is represented by the parameter $\delta$, defined in eq.~\eqref{delta}. In the effective theory, the stable and metastable minima are to the leading order located at $\Sigma=\pm v_0$ (and $\Pi_a=0$). A comparison of the values of the potentials gives then the matching condition $s_4v_0^3T=-\delta T^4$ which, using eq.~\eqref{v2T}, leads to the identification
\begin{equation}
s_4=-\frac\delta8=\frac{1}{2\pi^2}(\kappa_{-2}^--\kappa_{-2}^+).
\label{match3}
\end{equation}

The last parameter to be fixed, $s_5$, does not contribute to the energy difference of the two vacua, but does affect the shape of the effective potential. One simple and gauge invariant (although by no means unique) quantity sensitive to $s_5$ is the difference of the squared mass parameters at the two minima, i.e.~$m_\chi^2$ of eq.~\eqref{ZQCDcoeffs} and the analogous parameter at the metastable minimum. The corresponding quantities are also straightforward to evaluate in the full theory, see appendix~\ref{app:EQCD}, yielding the last perturbative matching condition,
\begin{equation}
2s_5-3s_4=\frac1{2\pi^2}(\kappa_0^+-\kappa_0^-).
\label{match4}
\end{equation}
It is interesting to note that the right hand side of this equation is proportional to the second derivative of the parameter $\delta$ with respect to the chemical potential(s), i.e.~the difference of the quark number susceptibilities in the stable and metastable vacua of the theory.


\section{Extended field configurations}
\label{sec:solitons}

Having now finished the leading order matching of ZQCD to the full theory, it is important to test its predictions in particular for quantities that are sensitive to the center symmetry. Perhaps the most straightforward such test is to study extended gauge field configurations, which probe the global structure of the effective potential. In the absence of fermions, and thus Z(2) breaking operators, we can construct a stable domain wall joining the two physically equivalent minima of the theory. With fermions, this is no longer possible, as the minima are not degenerate in energy, but one can still look for a rotationally invariant three-dimensional solution that represents a bubble of the stable vacuum in a metastable environment. Although this bubble evolves with time, its growth rate can be estimated using a semiclassical \emph{static} solution, representing a stationary point of a three-dimensional effective action with suitable boundary conditions~\cite{Coleman:1977py}.

Consider first a bubble wall configuration in the full theory, in which the (static) temporal gauge field $A_0$ depends only on the radial coordinate $r$ and points in the same direction in color space everywhere. It is described by a single scalar function $a(r)$, whose action has the form
\begin{equation}
S_{\text{eff}}=\beta\int_0^\infty\dd r\,4\pi r^2\left[\frac12\left(\frac{\dd a}{\dd r}\right)^2+V_{\text{eff}}(a)\right],
\label{bubble_action}
\end{equation}
where the potential $V_{\text{eff}}$ is obtainable from eq.~\eqref{EQCDpot}. One may then solve the equation of motion stemming from this action with the boundary condition $a(\infty)=2\pi T/g$, and use it to obtain the domain wall energy density and tension as well as the bubble profile, as has indeed be done in refs.~\cite{Bhattacharya:1990hk,Ignatius:1991nk} (see also ref.~\cite{Armoni:2010ny} for a recent similar calculation).

Within the effective theory, we first note that in order to minimize the energy cost of creating a bubble, it is clearly optimal to have the fields $\Sigma$, $\vec\Pi$ minimize the hard part of the potential~\eqref{VNLO} everywhere in space, i.e.~have them satisfy $\Sigma^2+\vec\Pi^2=v_0^2$. Recalling the identification of eq.~\eqref{v2T}, we see that we can express the fields in terms of one dimensionless function $\alpha$, ranging from 0 to 1, as $\Sigma=v_0\cos(\pi\alpha)$,  $\norvec\Pi=v_0\sin(\pi\alpha)$. Plugging these formulas into eq.~\eqref{VNLO}, we obtain upon a trivial shift the potential
\begin{equation}
\begin{split}
V_{\text{eff}}(\alpha)=&\frac{v_0^2}2(s_1-4s_3v_0^2)\sin^2(\pi\alpha)+\frac{v_0^4}4(s_2+4s_3)\sin^4(\pi\alpha)-\frac{v_0^3}{3\pi}|\sin(\pi\alpha)|^3+\\
&+\frac{v_0^3}2s_4\left[\cos^3(\pi\alpha)-1\right]+\frac{v_0^3}2s_5\cos(\pi\alpha)\sin^2(\pi\alpha),
\end{split}
\label{bubble_potential}
\end{equation}
using which the bubble profile can again be solved.\footnote{Inserting the values of the soft
parameters to the domain wall potential in the $N_{\rm f}=0$ case, one sees that the different terms
combine to
$V_{\text{eff}}(\alpha)=\frac43T^3\sin^2(\pi\alpha)\left[1-\frac1\pi|\sin(\pi\alpha)|\right]^2$,
closely reminiscent of the full theory domain wall potential (see in particular the discussion in
ref.~\cite{Altes:2008ic}). We would like to thank Chris Korthals Altes for pointing out this issue
to us.}

\begin{figure}
\begin{center}
\includegraphics[scale=1.25]{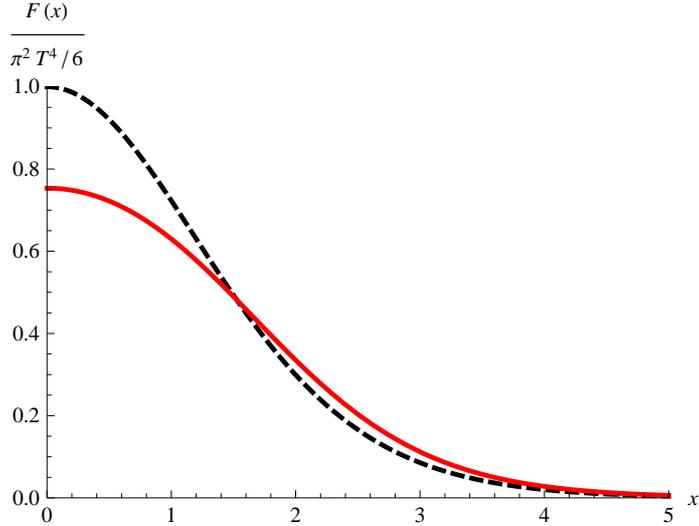}
\end{center}
\caption{Free energy profiles of the leading order domain wall solution in the center symmetric limit as a function of the dimensionless length variable $x\equiv gTr$, with the wall residing at $x=0$. The solid red curve corresponds to the prediction of the effective theory, while the dashed black one is the full Yang-Mills result. The boundary condition for this one-dimensional solution is $\alpha(-\infty)=0$ and $\alpha(+\infty)=1$.}
\label{fig:wall}
\end{figure}

Specializing for the moment to the domain wall calculation in the Z(2) invariant, pure Yang-Mills case, we observe that the potential of eq.~\eqref{bubble_potential} depends exactly on the two linear combinations of $s_{1,2,3}$ that were determined in our perturbative matching, cf.~eq.~\eqref{match2}. We conclude (in contrast to the claims of refs.~\cite{Vuorinen:2006nz,deForcrand:2008aw}) that the domain wall tension and profile become genuine predictions of the effective theory. Indeed, from eq.~\eqref{match2} we infer the results $s_1-4s_3v_0^2=2T/3$ and $s_2+4s_3=1/(3\pi^2T)$, using which we straightforwardly obtain as the domain wall tension
\begin{equation}
\sigma\approx4.899\times\frac{T^3}{g}\approx0.91\sigma_{\text{YM}},
\end{equation}
where $\sigma_{\text{YM}}=\left(\frac{2}{3}\right)^{3/2}\frac{\pi^2 T^3}{g}$ denotes the full theory result~\cite{Bhattacharya:1990hk}. In figure~\ref{fig:wall}, we plot the full and effective theory domain wall profiles, which we find to agree at a satisfactory level. These curves should be contrasted with those in figure~1 of ref.~\cite{deForcrand:2008aw}, where it was erroneously argued that one can fit the domain wall tension and width to the full theory values, based on the algebraic error discussed in section~\ref{subsec:couplings}.

Let us finally return to the three-dimensional bubble solution, relevant when dynamical quarks are present. Its formation and profile are determined by a balance between a volume energy gain, scaling like $\delta R^3$ (with the energy density difference $\delta$ introduced in eq.~\eqref{delta}), and a surface energy cost, scaling like $\sigma R^2$, where $R$ stands for the bubble radius in units of $1/(gT)$. We will not attempt a full numerical solution of the corresponding equation of motion, which is straightforward but not particularly illuminating, but instead provide an analytic approximation valid in the limit of parametrically small $\delta$, corresponding to weak Z(2) breaking effects. As we expect $R$ to scale like $1/\delta$, the explicit $r$-dependence of the action becomes then negligible and the bubble profile calculation reduces to the type of domain wall problem encountered above. This is usually called the thin wall approximation~\cite{Kapusta:2006pm}.

In the thin wall approximation, the bubble solution is universal in the sense that the bubble action (and therefore the radius) indeed only depend on the surface tension, obtained from the one-dimensional domain wall problem, and the energy density splitting of the two vacua. The critical radius of the bubble, obtained by \emph{maximizing} the action with respect to $R$, and the value of the action become
\begin{equation}
R_c=\frac{2}{\delta}\times\frac{\sigma}{T^3/g},\qquad S_{\text{bubble}}=\frac{16\pi}{3g^3\delta^2}\times\left(\frac{\sigma}{T^3/g}\right)^3.
\end{equation}
Quantitatively, the applicability of the thin wall approximation is determined by the condition that the radius of the bubble is much larger than the width of the domain wall, a quantity of order one in the dimensionless units. Using the full theory value for the surface tension, this translates to $\delta\lesssim 1$, which is certainly satisfied for quarks with $m_j\gtrsim T,\mu_j$, as $\delta$ is then exponentially suppressed.


\section{Conclusions and outlook}
\label{sec:conclusions}

In the paper at hand, we have constructed a dimensionally reduced effective theory for two-color QCD, generalizing the earlier works of refs.~\cite{Vuorinen:2006nz,deForcrand:2008aw} by including in the consideration the effects of fundamental quarks of in principle arbitrary masses and chemical potentials on the dynamics of the Wilson line. In the process, we also identified and corrected an important algebraic error in the original papers, which led to qualitatively new understanding of the structure of center symmetric effective theories. In particular, we found that the tension and profile of a domain wall separating the two Z(2) minima cannot be used as matching conditions between the effective and full theories, but that instead become genuine predictions of ZQCD.

The effective theory we constructed is formulated in terms of a `coarse grained Wilson line' type variable $\Z$ defined by eq.~\eqref{Zdef}, as well as the Lagrangian of eqs.~\eqref{lageff2}--\eqref{pot}. The theory has by construction a notion of the center symmetry transformation of the full theory, and is invariant under it in the absence of dynamical quarks. The matching of the effective theory to the full one was performed by requiring that the former reproduce the correct long distance physics of the latter, a task most conveniently accomplished by demanding that the theory reduces to EQCD upon integration out of the momentum scale $\pi T$. This we carried out explicitly in section~\ref{sec:matching}, where the $s_i$ parameters appearing in the effective theory Lagrangian were determined with the exception of one, for which a higher order computation is needed.

Upon fixing its soft parameters, our effective theory becomes fully predictive, as the physics of the distance scales $1/(gT)$ and larger is to a very good accuracy independent of the values of the hard parameters $h_i$ \cite{deForcrand:2008aw}. Thus, our theory is immediately amenable to nonperturbative lattice simulations, with which one may study two-color QCD over an extensive range of temperatures, quark chemical potentials and masses.\footnote{One particularly interesting direction to pursue in this respect is the continuation of the phase diagram of the theory to imaginary chemical potentials, where our effective theory should be able to describe the Roberge--Weiss transition (see also Refs.~\cite{Bonati:2010gi,deForcrand:2010he}). We are grateful to Ph.~de Forcrand for suggesting this possibility.} To this end, let us note three encouraging facts: First, as shown in refs.~\cite{Vuorinen:2003fs,Ipp:2006ij}, dimensional reduction does not as such require the temperature to be the largest energy scale in the system, but rather only that a modest scale hierarchy exist between the scale $\pi T$ and the electric screening mass, translating at high densities to a parametric condition $g\mu\lesssim\pi T$. Second, there is a priori no reason, why our effective theory could not be extended all the way to the limit of massless quarks, although it then has little relevance for the phase transition itself, which is driven by chiral dynamics. And third, it should be recalled that the only reason we have in this paper chosen to study two-color QCD and not the physical case of three colors is notational and computational simplicity. If the predictions of the effective theory turn out to match two-color lattice data well, then an obvious topic for further investigations will be the generalization of the present work to full three-color QCD.

Finally, another interesting topic for future work would clearly be to consider the relation of our effective theory to the strong coupling effective actions derived in refs.~\cite{Langelage:2008dj,Langelage:2010yr,Lottini:2011bj,Fromm:2011qi}. It appears that these two approaches are strongly complementary in the sense that they approach the deconfinement transition from opposite directions; whether this can be used to gain more insight into the dynamics of the transition itself remains to be seen.


\acknowledgments

We would like to thank Philippe de Forcrand, Chris Korthals Altes, and Owe Philipsen for carefully
reading the manuscript and suggesting numerous improvements, and J\"urgen Berges, Kenji
Fukushima, Mikko Laine, and Lorenz von Smekal for useful discussions. The work of T.Z.~was supported
by the German Academic Exchange Service (DAAD) and by the Helmholtz Graduate School for Hadron and
Ion Research. T.B.~and A.V.~were supported by the Sofja Kovalevskaja program of the Alexander von
Humboldt foundation. A.K.~was supported in part by the Natural Sciences and Engineering Research
Council of Canada and the Institute of Particle Physics (Canada).

\appendix


\section{EQCD parameters in the presence of massive fermions}
\label{app:EQCD}

The one-loop QCD effective potential evaluated in a static background $A_0$ field was first determined in refs.~\cite{Gross:1980br,Weiss:1980rj,Weiss:1981ev}. Rewriting it in a form invariant under \emph{global} SU(2) symmetry and parameterizing the gauge field as $A_0=\vec a\cdot\vec\sigma/2$, the result becomes
\begin{equation}
\begin{split}
V_{\text{eff}}(\vec a)=&\frac43\pi^2T^4\left\langle\frac{g|\vec a|}{2\pi T}\right\rangle^2\left(1-\left\langle\frac{g|\vec a|}{2\pi T}\right\rangle\right)^2-\\
&-2T\sum_{j=1}^{\Nf}\sum_\pm\int\frac{\dd^3\vek k}{(2\pi)^3}\log\left[1+2e^{-\beta(\epsilon_{j\vek k}\pm\mu_j)}\cos\frac{g|\vec a|}{2T}+e^{-2\beta(\epsilon_{j\vek k}\pm\mu_j)}\right],
\end{split}
\label{match_V}
\end{equation}
where $\langle\cdot\rangle$ denotes the fractional part of a real number ($\langle x\rangle=x-\lfloor x\rfloor$), $\mu_j$ the set of (flavor) quark number chemical potentials, and $\epsilon_{j\vek k}=\sqrt{\vek k^2+m_j^2}$ the dispersion relation of the $j$th quark flavor. Also, we used the shorthand notation $|\vec a|=\sqrt{\vec a\cdot\vec a}$. Expanding this expression in powers of $\vec a$ around zero and subtracting the contribution of the static modes (amounting to the term cubic in $\vec a$), one may readily identify the EQCD parameters
\begin{equation}
\begin{split}
m_\chi^2&=\frac{2g^2T^2}3-2g^2\sum_{j=1}^{\Nf}\int\frac{\dd^3\vek k}{(2\pi)^3}\bar f'(\epsilon_{j\vek k},\mu_j),\\
\tilde\lambda&=\frac{2g^4T}{3\pi^2}+\frac{g^4}6\sum_{j=1}^{\Nf}\int\frac{\dd^3\vek k}{(2\pi)^3}\bar f'''(\epsilon_{j\vek k},\mu_j),
\end{split}
\label{match_coeff}
\end{equation}
where $\bar f(x,\mu)=[f(x+\mu)+f(x-\mu)]/2$, the prime denotes differentiation with respect to $x$, and $f(x)=1/(e^{\beta x}+1)$ is the Fermi--Dirac distribution function.

Compact as the above expressions~\eqref{match_V} and~\eqref{match_coeff} are, one can further evaluate the integrals over the quark momentum analytically in terms of the modified Bessel function of the second kind, $K_n$. Expanding the logarithm in powers of fugacity and using some identities for the Bessel functions, one obtains the result
\begin{equation}
\begin{split}
V_{\text{eff}}(\vec a)=&\frac43\pi^2T^4\left\langle\frac{g|\vec a|}{2\pi T}\right\rangle^2\left(1-\left\langle\frac{g|\vec a|}{2\pi T}\right\rangle\right)^2+\\
&+\frac{4T^2}{\pi^2}\sum_{j=1}^{\Nf}m_j^2\sum_{n=1}^\infty\frac{(-1)^{n}}{n^2}K_2(n\beta m_j)\cosh(n\beta\mu_j)\cos\frac{ng|\vec a|}{2T},
\end{split}
\label{EQCDpot}
\end{equation}
where the sum converges as long as $\mu_j<m_j$ for all quark flavors. Analogously, one derives by differentiation analytic expressions for the EQCD mass parameter and quartic coupling,
\begin{equation}
\begin{split}
m_\chi^2&=\frac{2g^2T^2}3-\frac{g^2}{\pi^2}\sum_{j=1}^{\Nf}m_j^2\sum_{n=1}^\infty(-1)^nK_2(n\beta m_j)\cosh(n\beta\mu_j),\\
\tilde\lambda&=\frac{2g^4T}{3\pi^2}+\frac{g^4}{12\pi^2T}\sum_{j=1}^{\Nf}m_j^2\sum_{n=1}^\infty(-1)^nn^2K_2(n\beta m_j)\cosh(n\beta\mu_j).
\label{EQCDcoeffs}
\end{split}
\end{equation}
When the quark mass is parametrically larger than both the temperature and the respective chemical potential, the infinite series in eq.~\eqref{EQCDcoeffs} can be replaced by its asymptotic form,
\begin{equation}
\begin{split}
m_\chi^2&\approx\frac{2g^2T^2}3+2g^2T^2\sum_{j=1}^{\Nf}\left(\frac{m_j}{2\pi T}\right)^{3/2}{\rm e}^{-m_j/T}\cosh(\beta\mu_j),\\
\tilde\lambda&\approx\frac{2g^4T}{3\pi^2}-\frac{g^4T}6\sum_{j=1}^{\Nf}\left(\frac{m_j}{2\pi T}\right)^{3/2}{\rm e}^{-m_j/T}\cosh(\beta\mu_j).
\end{split}
\label{EQCDcoeffsasympt}
\end{equation}
On the other hand, for massless quarks at vanishing chemical potentials, the integrals in eq.~\eqref{match_coeff} are readily evaluated analytically and one finds $m_\chi^2=(2g^2T^2/3)[1+(\Nf/4)]$ and $\tilde\lambda=(2g^4T/3\pi^2)[1-(\Nf/8)]$, in agreement with ref.~\cite{Kajantie:1997tt}.

As the above infinite series containing Bessel and hyperbolic functions will appear frequently in our results, it is convenient to introduce a shorthand notation,
\begin{equation}
\kappa_\ell^\pm=\sum_{j=1}^{\Nf}(\beta m_j)^2\sum_{n=1}^\infty(\pm1)^n n^\ell K_2(n\beta m_j)\cosh(n\beta\mu_j),
\label{kappal}
\end{equation}
in terms of which the EQCD parameters~\eqref{EQCDcoeffs} take the simple forms $m_\chi^2=(2g^2T^2/3)[1-(3\kappa_0^-/2\pi^2)]$ and $\tilde\lambda=(2g^4T/3\pi^2)[1+(\kappa_2^-/8)]$.\footnote{Note that at any given time, the infinite sum can be replaced by the corresponding integral expression. This is in particular necessary for reasons of convergence, if $m_j<\mu_j$ for some quark flavor.} To introduce one final piece of notation, observe that in the presence of dynamical quarks, the potential of eq.~\eqref{EQCDpot} has only one global minimum (up to periodicity) at $\norvec a=0$, while the point $\norvec a=2\pi T/g$ corresponds to a mere local minimum. The most important quantity carrying information on the explicit Z(2) breaking due to dynamical quarks is thus the energy density difference between the two minima. It can be encoded in a single dimensionless parameter
\begin{equation}
\delta\equiv\frac{V_{\text{eff}}(g\norvec a=2\pi T)-V_{\text{eff}}(g\norvec a=0)}{T^4}=\frac4{\pi^2}(\kappa_{-2}^+-\kappa_{-2}^-).
\label{delta}
\end{equation}


\section{Center symmetry for the SU(2) gauge group}
\label{app:center}

The zero component of the gauge field $A_0(\tau)$ (dependence on the spatial coordinates does not play any role in what follows and is thus suppressed) transforms under the local transformation $s(\tau)\in$ SU(2) as $\tilde A_0(\tau)=s(\tau)A_0(\tau)s(\tau)^\dag+\frac\imag gs(\tau)\partial_\tau s(\tau)^\dag$. Under the same transformation, the (untraced) Wilson line operator defined in eq.~\eqref{omega} transforms as $\tilde\Omega=s(\beta)\Omega s(0)^\dag$. Let us now write the gauge field as $A_0(\tau)=\vec a(\tau)\cdot\vec\sigma/2$, and the most general SU(2) gauge transformation as $s(\tau)=\exp[\imag\varphi(\tau)\vec n(\tau)\cdot\vec\sigma]$, where $\vec n(\tau)$ is a unit vector. In this representation, the gauge field transforms as
\begin{equation}
\begin{split}
\tilde{\vec a}=&\vec n(\vec n\cdot\vec a)+[\vec a-\vec n(\vec n\cdot\vec a)]\cos2\varphi-(\vec n\times\vec a)\sin2\varphi+\\
&+\frac1g\left[2\vec n\varphi'+\vec n'\sin2\varphi-2(\vec n\times\vec n')\sin^2\varphi\right],
\end{split}
\end{equation}
where the prime denotes a derivative with respect to $\tau$. Demanding that the gauge transformation preserves the periodicity of the gauge field, $\vec a(\beta)=\vec a(0)$, leads to the conditions
\begin{equation}
\varphi(\beta)=\varphi(0)+N\pi,\qquad \varphi'(\beta)=\varphi'(0),\qquad \vec n(\beta)=\vec n(0),\qquad \vec n'(\beta)=\vec n'(0),
\end{equation}
up to an overall minus sign, which only matters if we require the parameters $\varphi,\vec n$ to change continuously with $\tau$. In either case, the unitary matrix $s(\tau)$ satisfies $s(\beta)=(-1)^Ns(0)$ for some integer $N$, which is precisely a transformation of the Z(2) center of the gauge group.

Let us now specialize to the Polyakov gauge, in which $A_0$ is diagonal and independent of $\tau$. Which gauge transformations from the local SU(2) group preserve this structure? Obviously, $\vec n$ must point in the third direction at all times. Preservation of time independence of $\vec a$ then results in the condition $\varphi'=\text{const}$, and the admissible gauge transformations take the form $\vec n=(0,0,1)$ and $\varphi(\tau)=\varphi(0)+N\pi\tau/\beta$. As a consequence, the gauge field transforms by a mere overall shift, $a_3\to a_3+2N\pi T/g$. It is worth emphasizing, though, that this conclusion only holds in the Polyakov gauge, as otherwise the \emph{vector function} $\vec a(\tau)$ transforms in a rather complicated manner. In any case, since the nontrivial center transformations correspond to time-dependent $s(\tau)$, while there is no time in the three-dimensional effective theory, this effective theory must be augmented with a suitable \emph{definition} of the center symmetry.


\section{One-loop effective potential of ZQCD}
\label{app:effpot}

The derivation of the one-loop effective potential of the theory defined by eq.~\eqref{lageff2} follows closely appendix~A of ref.~\cite{deForcrand:2008aw}, and we will therefore merely write down the result here. In practice, we choose $\vec\Pi$ to point in the third color direction, $\Pi_a=\norvec\Pi\delta_{a3}$, relying on the SU(2) invariance of the theory. The effective potential in a general $R_\xi$  renormalizable gauge then consists of the tree-level contribution, the gluon and ghost loops, the loop in the mixed $\Sigma\Pi_3$ sector, as well as a separate contribution from the $\Pi_{1,2}$ loops,
\begin{equation}
\begin{split}
V_{\text{eff}}&=\frac1{g_3^2}V_{\text{tree}}+V_{\vek A+\text{gh}}+V_{\Sigma\Pi_3}+V_{\Pi_{1,2}},\\
V_{\text{tree}}&=b_1\Sigma^2+b_2\vec\Pi^2+c_1\Sigma^4+c_2(\vec\Pi^2)^2+c_3\Sigma^2\vec\Pi^2+d_1\Sigma^3+d_2\Sigma\vec\Pi^2,\\
V_{\vek A+\text{gh}}&=-\frac{\norvec\Pi^3}{6\pi}\bigl(2-\xi^{3/2}\bigr),\qquad
V_{\Sigma\Pi_3}=-\frac1{12\pi}\left[(m_+^2)^{3/2}+(m_-^2)^{3/2}\right],\\
V_{\Pi_{1,2}}&=-\frac1{6\pi}(2b_2+4c_2\vec\Pi^2+2c_3\Sigma^2+2d_2\Sigma+\xi\vec\Pi^2)^{3/2},
\end{split}
\label{Veff1loop}
\end{equation}
where
\begin{equation}
\begin{split}
m_\pm^2=&b_1+b_2+6c_1\Sigma^2+6c_2\vec\Pi^2+c_3(\Sigma^2+\vec\Pi^2)+(3d_1+d_2)\Sigma\pm\\
&\pm\sqrt{\left[b_1-b_2+6c_1\Sigma^2-6c_2\vec\Pi^2+c_3(\vec\Pi^2-\Sigma^2)+(3d_1-d_2)\Sigma\right]^2+4\vec\Pi^2(2c_3\Sigma+d_2)^2}.
\end{split}
\end{equation}
Using the parametrization of the couplings~\eqref{bd1} and assuming that the background fields $\Sigma,\vec\Pi$ are of order $\mathcal O(g_3^0)$ or smaller, we immediately obtain an expansion for the effective potential to next-to-leading order in the coupling,
\begin{equation}
\begin{split}
V_{\text{eff}}\approx&\frac1{g_3^2}\left[\frac{h_1}2(\Sigma^2+\vec\Pi^2)+\frac{h_2}4(\Sigma^2+\vec\Pi^2)^2\right]+\frac{s_1}2\vec\Pi^2+\frac{s_2}4(\vec\Pi^2)^2+s_3\Sigma^4+\frac{s_4}2\Sigma^3+\frac{s_5}2\Sigma\vec\Pi^2-\\
&-\frac{\norvec\Pi^3}{6\pi}\bigl(2-\xi^{3/2}\bigr)-\frac1{6\pi}(\tilde m_-^2+\xi\vec\Pi^2)^{3/2}-\frac1{12\pi}(\tilde m_+^3+\tilde m_-^3),
\end{split}
\label{VNLO}
\end{equation}
where $\tilde m_+^2=h_1+3h_2(\Sigma^2+\vec\Pi^2)$ and $\tilde m_-^2=h_1+h_2(\Sigma^2+\vec\Pi^2)$.


\bibliographystyle{JHEP}
\bibliography{refs}

\end{document}